\documentclass[sigconf]{acmart}
\AtBeginDocument{%
  \providecommand\BibTeX{{%
    \normalfont B\kern-0.5em{\scshape i\kern-0.25em b}\kern-0.8em\TeX}}}

\usepackage{multirow}
\usepackage{subcaption}
\usepackage{tikz}
\usepackage{pgfplots}
\pgfplotsset{compat=newest}
\pgfplotsset{plot coordinates/math parser=false}
\newlength\fheight
\newlength\fwidth
\usetikzlibrary{plotmarks,decorations,backgrounds,calc,spy}
\usetikzlibrary{decorations.markings}
\usepgfplotslibrary{patchplots,groupplots}
\usepackage{tikzscale}
\usepackage{graphicx}

\hypersetup{draft}

\copyrightyear{2023}
\acmYear{2023}
\setcopyright{acmlicensed}\acmConference[WNS3 2023]{2023 Workshop on ns-3}{June 28--29, 2023}{Arlington, VA, USA}
\acmBooktitle{2023 Workshop on ns-3 (WNS3 2023), June 28--29, 2023, Arlington, VA, USA}
\acmPrice{15.00}
\acmDOI{10.1145/3592149.3592158}
\acmISBN{979-8-4007-0747-6/23/06}

\usepackage{glossaries}

\newacronym{ntn}{NTN}{non-terrestrial network}
\newacronym{3gpp}{3GPP}{3rd Generation Partnership Project}
\newacronym{5g}{5G}{5th generation}
\newacronym{6g}{6G}{6th generation}
\newacronym{m2m}{M2M}{Machine-to-Machine}
\newacronym{v2x}{V2X}{Vehicle-to-Everything}
\newacronym{nr}{NR}{New Radio}
\newacronym{pl}{PL}{path loss}
\newacronym{geo}{GEO}{Geostationary Equatorial Orbit}
\newacronym{meo}{MEO}{medium earth orbit}
\newacronym{leo}{LEO}{Low Earth Orbit}
\newacronym{upa}{UPA}{uniform planar array}
\newacronym{vsat}{VSAT}{very small aperture terminal}
\newacronym{nb}{NB}{node base}
\newacronym{ue}{UE}{user equipment}
\newacronym{mmwave}{mmWave}{millimeter wave}
\newacronym{mimo}{MIMO}{multiple input multiple output}
\newacronym{los}{LOS}{line-of-sight}
\newacronym{nlos}{NLOS}{non-line-of-sight}
\newacronym{hap}{HAP}{High Altitute Platform}
\newacronym{iot}{IoT}{Internet of Things}
\newacronym{mtc}{MTC}{machine type communications}
\newacronym{miot}{mIoT}{massive internet of things}
\newacronym{gcs}{GCS}{global coordinate system}
\newacronym{lcs}{LCS}{local coordinate system}
\newacronym{uav}{UAV}{Unmanned Aerial Vehicle}
\newacronym{ns3}{ns-3}{network simulator 3}
\newacronym{nfv}{NFV}{network function virtualization}
\newacronym{sdn}{SDN}{software defined networking}
\newacronym{isl}{ISL}{inter-satellite link}
\newacronym{ul}{UL}{uplink}
\newacronym{dl}{DL}{downlink}
\newacronym{ecef}{ECEF}{Earth-Centered Earth-Fixed}
\newacronym{sf}{SF}{shadow fading}
\newacronym{enu}{ENU}{east north up}
\newacronym{snr}{SNR}{signal to noise ratio}
\newacronym{mec}{MEC}{mobile edge cloud}
\newacronym{hpbw}{HPBW}{half power beam width}
\newacronym{rb}{RB}{resource block}
\newacronym{udp}{UDP}{user datagram protocol}
\newacronym{fspl}{FSPL}{free space path loss}
\newacronym{cnr}{CNR}{carrier-to-noise ratio}

\begin{document}

\title{Implementation of a Channel Model for Non-Terrestrial Networks in ns-3}

\author{Mattia Sandri, Matteo Pagin, Marco Giordani, and Michele Zorzi}
\affiliation{%
  \institution{Department of Information Engineering, University of Padova, Padova, Italy}
\country{}
}
\email{{sandrimatt, paginmatte, giordani, zorzi} @dei.unipd.it}

\begin{abstract}
    While the \gls{5g} of mobile networks has landed in the commercial area, the research community is exploring new functionalities for \gls{6g} networks, for example \glspl{ntn} via space/air nodes such as \glspl{uav}, \glspl{hap} or satellites.  Specifically, satellite-based communication offers new opportunities for future wireless applications, such as providing connectivity to remote or otherwise unconnected areas, complementing terrestrial networks to reduce connection downtime, as well as increasing traffic efficiency in hot spot areas. 
    In this context, an accurate characterization of the NTN channel is the first step towards proper protocol design.  
    Along these lines, this paper provides an ns-3 implementation of the \gls{3gpp} channel and antenna models for \gls{ntn} described in Technical Report 38.811. %
    In particular, we extend the ns-3 code base with new modules to implement the attenuation of the signal in air/space due to atmospheric gases and scintillation, and new mobility and fading models to account for the Geocentric Cartesian coordinate system of satellites. Finally, we validate the accuracy of our ns-3 module via simulations against 3GPP calibration results.
\end{abstract}



\maketitle
\glsresetall

\section{Introduction}
The large-scale deployment of cellular networks started at the end of last century.
Since then, the number of connected users has rapidly increased, together with the capabilities provided by wireless networks, now towards their \gls{6g}~\cite{giordani2020toward}.
This, together with the introduction of new applications in which millions or even billions of devices require connectivity services, such as \gls{iot} and \gls{v2x}, poses new challenges in terms of system capacity, network coverage, and service reliability. 
In this context, the research community is investigating the adoption of \glspl{ntn}~\cite{giordani2021non}, and the \gls{3gpp} has consolidated the possible use of \gls{ntn} into the \gls{nr} standard in Release~17~\cite{21917}.
In \gls{ntn}, \glspl{uav}, \glspl{hap}, and satellites can offer new connectivity opportunities, e.g., complementing already existing cellular systems, or providing broadband coverage to rural regions where it would otherwise be infeasible to install cellular towers~\cite{Chaoub20216g}.

Satellites have been used from the 1990s to provide basic services such as phone and Internet access. Notably,  \gls{geo} satellites orbit at 35\,786 km, and offer global coverage at limited costs, despite the very large propagation delays. 
Only in the early 2010s have the costs for satellite launch and maintenance become low enough to allow for very large constellations of satellites to be launched in the \gls{leo}~\cite{satcost}. In this case, LEO provides wide coverage of the Earth, while promoting low~latency. 

Besides satellites, both \glspl{hap} and \glspl{uav} stand out as valid cost-effective alternatives for NTN. 
\glspl{uav}, flying at low altitudes (typically no more than 1 km), can guarantee on-demand support for ground networks, for example providing immediate assistance when cellular towers are overwhelmed or unavailable. \glspl{hap} operate in the stratosphere (from 20 to 50 km), and can be used to shape large coverage beams in  unpopulated areas, or provide services like backhauling and \gls{mec}, e.g., to gather and process data generated on the ground~\cite{traspadini2023real,wang2020potential}. %

Despite these premises, however, communication using space or airborne vehicles introduces new challenges compared to a terrestrial base station, including (i) severe \gls{pl} due to the longer propagation distance; (ii) additional attenuation from the atmosphere, such as scintillation, rain and clouds; (iii) Doppler shift due to the orbital mobility of satellites; and (iv) additional delays, mainly for propagation.
While experiments with real testbeds are impractical due to limitations in the scalability and flexibility of platforms, as well as the high cost of hardware components, the option to test network configurations via simulations in a sandbox environment facilitates the research process. Furthermore, an open-source simulator encourages research in the field, and offers industries and research institutions a better way to categorize and evaluate technologies. Among other simulators, ns-3 is a popular and effective open-source highly-customizable software to evaluate the end-to-end full-stack performance of networks~\cite{henderson2008network}.

In this context, most simulators for \gls{ntn}, e.g., 5G K-Simulator \cite{5gksimulator}, 5GVienna~\cite{vienna5gsimulator}, or Simu5G \cite{simu5g}, are proprietary, or require some type of commercial license for usage. Some others, while being open-source, sacrifice the accuracy of the higher layers to reduce the computational complexity, e.g., 5G-air-simulator \cite{5gairsimulator}. 

To fill these gaps, in this paper we present a new open-source module for ns-3 that implements the \gls{ntn} channel model based on the 3GPP specifications described in Technical Report 38.811~\cite{38811}.\footnote{The source code is publicly available, and can be found at \cite{ntngitlab}. Notice that the authors are working with the ns-3 developers to merge the proposed NTN module into ns-3-dev, to promote future developments.}
While our implementation is mainly related to the channel and the physical layer, a deep understanding of the propagation model is the first step towards proper protocol design~\cite{lecci2021accuracy}, which makes our module a valuable and accurate tool in the study of NTN.
Specifically, our module introduces: (i) new simulation scenarios for NTN; (ii) a new \gls{pl} model for the air/space channel in a wide range of frequencies (from 0.5 GHz to 100 GHz); (iii) the characterization of atmospheric absorption; (iv) a new fast fading model for the space environment; (v) an antenna model for both terrestrial and non-terrestrial nodes; and (vi) a new coordinate system to account for the Geocentric Cartesian coordinate system of satellites. 

The rest of the paper is organized as follows. In Section~\ref{sec:channel} we describe the 3GPP channel for NTN, while our ns-3 implementation is reported in Section~\ref{sec:implementation}. In Section~\ref{sec:results} we validate the accuracy of our module against the 3GPP calibration results reported in 38.821~\cite{38821}, and provide numerical results to measure the link-level and end-to-end performance as a function of different NTN parameters. Finally, Section~\ref{sec:conclusions} concludes the paper with suggestions for future research.

\section{The 3GPP Channel Model for Non-Terrestrial Networks}
\label{sec:channel}
The channel model for \gls{ntn} has been formalized by the \gls{3gpp} in TR 38.811~\cite{38811}, and is based on the structure of the cellular channel model described in TR 38.901 \cite{38901}, already implemented in ns-3~\cite{mmwavemodulens3}. 
Following this rationale, in the following sections we describe the \gls{ntn} channel model, including path loss, absorption, fading, and antenna models, as well as the coordinate system for satellites.

\subsection{Scenarios and Path Loss Condition}
Similarly to the cellular channel model in \cite{38901}, the \gls{ntn} model gives the option to run simulations in four {scenarios}, to represent different propagation environments. Specifically:
\begin{itemize}
    \item Dense Urban: Extremely dense environment, with tall buildings acting as potential blockers.
    \item Urban: City environment, with buildings.
    \item Suburban: Small city, with up to two-storey buildings.
    \item Rural: Open field environment, with little or no buildings.
\end{itemize}
For satellites, only outdoor communication is possible, since attenuation from buildings would be enough to make the signal unusable. For \glspl{hap} or \glspl{uav}, instead, indoor communication is feasible, even though not yet implemented in our module. 

The 3GPP defines both \gls{los} and \gls{nlos} propagation, where the probability depends on the scenario and the elevation angle. The latter is defined as the angle between the horizon plane of the ground terminal and the vector pointing to the NTN platform. 

\subsection{Path Loss}
The basic path loss (in dB) can be written as:
\begin{equation}
    PL_{b} = FSPL(d,f_{c})+SF+CL(\alpha,f_{c}).
    \label{eq:pl}
\end{equation}
The first term is the free space path loss, which can be calculated as:
\begin{equation}
    FSPL(d,f_{c}) = 32.45 + 20\log_{10}(f_{c}) + 20\log_{10}(d),
\end{equation}
where $f_{c}$ is the carrier frequency in GHz and $d$ is the distance between the transmitter and the receiver in meters.
Notice that, while the channel model is valid for frequencies from 0.5 GHz to 100 GHz, two frequency bands are targeted in NTN, i.e., the S-band for frequencies below 6 GHz and the Ka-band for frequencies around 20 (30) GHz for downlink (uplink) transmissions, i.e., in the millimeter-wave spectrum~\cite{giordani2020satellite}.

In Equation~\eqref{eq:pl}, $SF$ represents the \gls{sf}, and is modeled as a log-normal random variable, i.e., $SF \sim N\left ( 0,\sigma_{SF}^{2} \right )$. In order to calculate this variance, the model requires four parameters: the type of scenario, the frequency, the path loss condition (\gls{los} or \gls{nlos}), and the elevation angle. These parameters are used to find the correct entry in a table, given in~\cite{38811}.
A similar process is needed to calculate the clutter loss $CL(\alpha,f_{c})$.

\subsection{Atmospheric Absorption}
\label{sub:atm}
The attenuation introduced by the presence of atmospheric gasses is a marginal factor in the terrestrial channel.
This is not the case for the \gls{ntn} channel, where atmospheric absorption plays a crucial role in the overall link budget. 
A complete and accurate characterization of the atmospheric attenuation is given in the ITU model~ \cite{itup676}, and depends on a set of parameters which is usually difficult to retrieve in simulations, such as absolute humidity, dry air pressure, water-vapor density and water-vapor partial pressure. 
Therefore, the \gls{3gpp} offers a simplified method considering only ground users placed at the sea level, with an elevation angle fixed to 90 degrees, and considering mean annual global values for the rest of the parameters. 
For elevation angles different than 90 degrees, given the zenith attenuation $A_{zenith}(f_{c})$, the additional path loss due to atmospheric gasses is computed as
\begin{equation}
    PL_{A,dB}\left ( \alpha,f_{c} \right )=\frac{A_{zenith}\left ( f_{c} \right ) }{\sin(\alpha)},
\end{equation}
where $\alpha$ is the actual elevation angle.
Atmospheric absorption should be considered only for frequencies above 10 GHz, or for any frequency in case of $\alpha<10$ degrees.

\subsection{Scintillation}
Scintillation corresponds to the rapid fluctuation in amplitude and phase of the received signal, caused by the variation of the refractive index of the channel. 
Specifically, scintillation depends on location, time of the day, season, and solar and geomagnetic activity. Stronger levels of scintillation are observed only at high latitudes  (more than 60 degrees), in auroral and polar regions.

While a complete absorption model would unnecessarily complicate system-level simulations, the 3GPP recommends a simplified model for scintillation~\cite{38811}, which is structured into two components: ionospheric scintillation and tropospheric scintillation.

\subsubsection{Ionospheric Scintillation}
Ionospheric scintillation is modeled based on the Gigahertz Scintillation Model~\cite{itup531}. While for the purpose of system–level simulations ionospheric scintillation is generally negligible at mid latitudes (between 20 and 60 degrees) or at above-6 GHz frequencies, in all other latitudes and conditions it can be modeled as
\begin{equation}
    PL_{S,dB} = \left ( \frac{f_{c}}{4} \right )^{-1.5}\frac{P_{fluc}\left ( 4 \text{ GHz} \right )}{\sqrt{2}},
\end{equation}
where $f_{c}$ is the carrier frequency, and $P_{fluc}\left ( 4 \text{ GHz} \right )$ is a scaling factor representing the ionospheric attenuation level at 99\% of the time observed in Hong Kong between March 1977 and March 1978 at a frequency of 4 GHz~\cite[Figure~6.6.6.1.4-1]{38811}.

\subsubsection{Tropospheric Scintillation}
Unlike ionospheric scintillation, the effect of tropospheric scintillation increases with the frequency, and becomes significant above 10 GHz. Furthermore, it increases at low elevation due to the longer path of the signal.  %
In these conditions, tropospheric scintillation is due to sudden changes in the refractive index due to the variation of temperature, water vapor content, and barometric pressure. For system-level simulations, the additional attenuation due to tropospheric scintillation is modeled as the attenuation level at 99\% of the time observed in Tolouse at 20 GHz, reported in~\cite[Figure~6.6.6.2.1-1]{38811}. 

\subsection{Fast Fading}
As far as fading is concerned, the 3GPP introduces both a flat-fading and a frequency-selective model. 
The flat-fading model, however, can be applied only if specific conditions are met, including (i) minimum elevation angle of 20 degrees, (ii) quasi-LOS propagation, (iii) communication in the S-band, (iv) channel bandwidth of at most 5 MHz, and (v) rural, suburban or urban scenario.
Hence, we consider the more general, though complex, frequency-selective fading model for \gls{ntn}. Then, fading is based on the TR 38.901 model for cellular networks~\cite{38901} (already implemented in ns-3 in the \texttt{mmwave} module~\cite{mmwavemodulens3}), but with different parameters as described  in~\cite[Section~6.7.2]{38811}.

\subsection{Antenna Model}
Different antenna models are defined, depending on the device (satellite, HAP or UAV, and ground terminal). 
For satellites, the \gls{3gpp} suggests to use a circular aperture antenna model. Circular aperture antennas are reflector antennas that offer circular polarization. The normalized antenna gain pattern is given by
\begin{equation}
\label{eq:eq4}
    G(\theta) =
    \begin{cases}
    1  & \theta=0; \\
    4\left | \frac{J_{1}\left ( k\cdot \ell\cdot \sin\theta \right )}{k\cdot \ell\cdot \sin\theta} \right |^{2} & 0<\left | \theta \right |\leq 90^{\circ},
    \end{cases}
\end{equation}
where $J_{1}(\cdot)$ is the Bessel function of the first kind and first order, $\ell$ is the radius of the antenna's circular aperture and, given a carrier frequency  $f_{c}$, the value of $k$ is equal to $k={2\pi f_{c}}/{c}$, where $c$ is the speed of light in vacuum.

When considering flying vehicles that are not satellites, such as \glspl{hap} or \glspl{uav}, the lower distance makes it possible to use \gls{upa} antennas, that is the current standard for \glspl{ue} and eNB/gNB nodes in cellular networks according to the TR 38.901 model~\cite{38901}.

Finally, for terrestrial terminals, the 3GPP suggests to use either \gls{upa} antennas or \gls{vsat} antennas. The latter model, in particular, is a common choice in satellite communication, and consists of a circular reflector antenna of small size (less than 1 m of diameter), to be typically placed on roofs pointing at the sky. The \gls{vsat} radiation pattern is the same as that of circular aperture antennas for satellites, and is given in Equation~\eqref{eq:eq4}.

\subsection{Coordinate System}
\label{sec:coord_model}
In general, the 3GPP defines a simple Cartesian coordinate system where the position of each node is uniquely described by a set of three values, $(x,y,z)$, where $x$ and $y$ define the ground plane, and $z$ represents the height of the node. 
While this model is accurate enough to describe scenarios where nodes are deployed at close distance (e.g., a few hundred meters), it is not for scenarios where end nodes are placed hundreds (or thousands) of kilometers apart such as in the NTN environment. In this case, the Earth's curvature, as well as the elevation angle, play an important role.
Therefore, the \gls{3gpp} suggests to use a Geocentric Cartesian coordinate system (or \gls{ecef} system), where the position of a node is still described by three values $(x,y,z)$, but now the origin of the axes lays in the center of the Earth, which is modeled as a sphere of radius $R=6\,371$ km. 
The x-y plane defines the equatorial plane, with the x-axis pointing at 0-degree longitude, the y-axis pointing at 90-degree longitude, and the z-axis pointing at the geographical North Pole. 
Then, terrestrial nodes on the surface of the Earth are deployed so that $\sqrt{x^{2}+y^{2}+z^{2}}=R$, while aerial/space nodes flying/orbiting around the Earth are deployed so that $\sqrt{x^{2}+y^{2}+z^{2}}> R$.

\section{Implementation}
\label{sec:implementation}
\begin{figure*}
    \includegraphics[width=\textwidth]{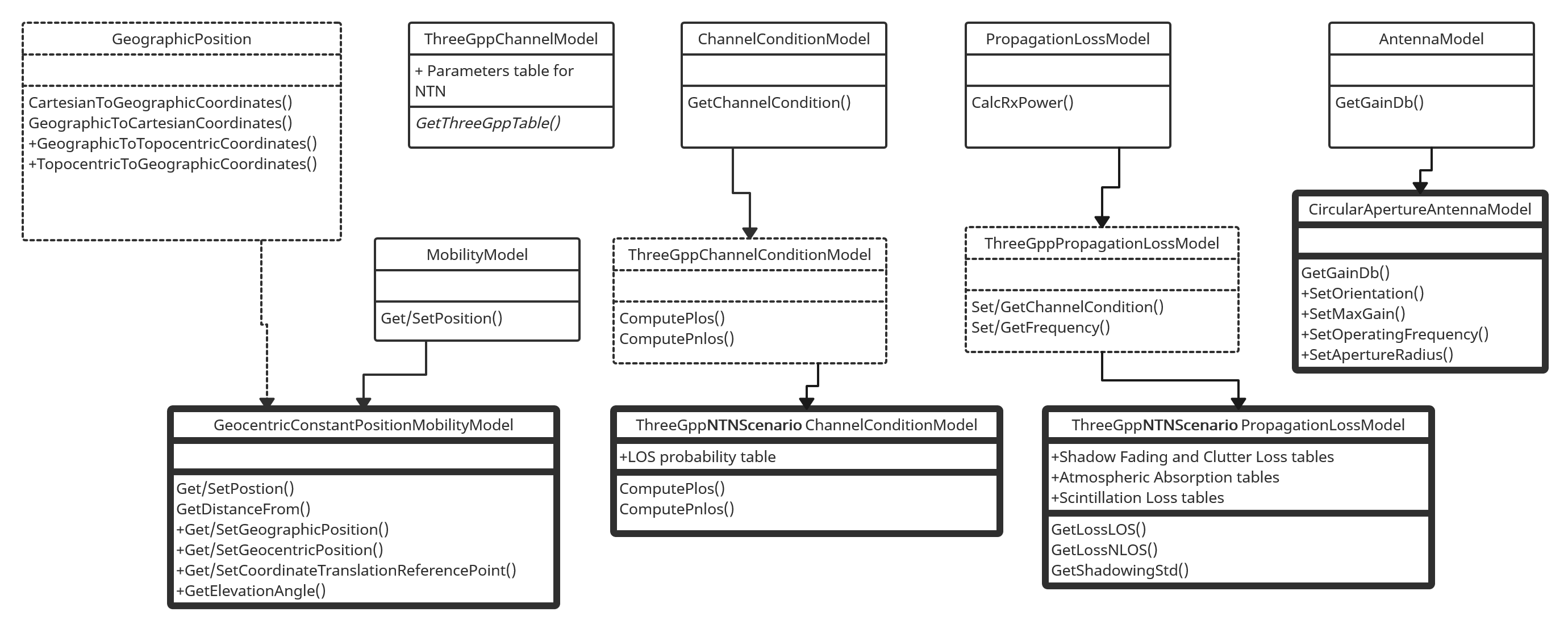}
    \caption{Simplified UML Diagram, which Depicts the most Significant Changes which we Introduced to ns-3. Bold Classes Represent the Newly Implemented Ones, while Dotted Classes Are Pre-existing Ones that Have Been Modified}
    \label{fig:ntn-uml}
\end{figure*}

The proposed ns-3 implementation of the 3GPP TR 38.811~\cite{38811} \gls{ntn} channel model is based upon the TR 38.901 model presented in~\cite{mmwavemodulens3}. Despite the fact that the newly introduced methods and classes are designed with the goal of introducing minimal changes to the existing ns-3 APIs, %
some modifications to the existing code structure are still required. A schematic of these changes, which we make publicly available\footnote{\url{https://gitlab.com/mattiasandri/ns-3-ntn/-/tree/ntn-dev}}, can be found in Figure~\ref{fig:ntn-uml}.

The remainder of this section describes more in detail our implementation of the NTN channel model of~\cite{38811} in ns-3. 

\subsection{Small-scale Fading}
The most significant modifications to the pre-existing ns-3 classes concern the \texttt{Three\-Gpp\-Channel\-Model} class, which computes the small-scale propagation phenomena in the form of a complex channel matrix. Indeed, conversely from the procedure implemented in \cite{mmwavemodulens3}, in the \gls{ntn} channel model of~\cite{38811} most channel parameters depend on all the propagation scenario, \gls{los} condition, carrier frequency and elevation angle variables.  
To account for this, we store the small-scale fading parameters in a \emph{nested map}. The choice of this data-structure is motivated by the good trade-off between code efficiency and readability it provides, considering that the number of possible combinations of the input parameters is particularly high, i.e., $144$.

In particular, we change the signature of the \texttt{GetThreeGppTable()} method to include the \texttt{Mobility\-Model} instances of both transmitting and receiving nodes. In such a way, we account for the dependence of the small scale parameters with respect to the elevation angle, which the model of~\cite{38811} exhibits.
Finally, we include the required angular scaling factors for the propagation scenarios that have a lower number of clusters than those described in TR 38.901~\cite{38901}.

\subsection{Coordinate Systems}
Instead of the coordinate system described in~\cite{38901}, the NTN channel model of~\cite{38811} considers the \emph{Geocentric Cartesian} (or \gls{ecef}) coordinate system.
We introduce this reference system in ns-3 via the \texttt{Geographic\-Positions} class, which provides methods to translate points represented using the coordinate system of~\cite{38901} to/from that of~\cite{38811}.

Moreover, with usability in mind, we also implement a geographic coordinate system which allows ns-3 users to specify positions using the system of~\cite{38811} in a more convenient manner. This auxiliary reference system represents positions as points exhibiting a relative altitude from their projection on the surface of the Earth. That is to say, any location on, or possibly above, the Earth is referenced by a longitude $\phi$, a latitude $\lambda$ and an altitude $h$. 
To this end, we implement in the \texttt{Geographic\-Positions} class the methods \texttt{Geographic\-To\-Topocentric\-Coordinates} and \texttt{Topocentric\-To\-Geographic\-Coordinates}, which can be used to translate positions between geocentric and non-geocentric geographic coordinate systems.

For the conversion between any of the newly introduced models, and the cartesian reference system of~\cite{38901}, we introduce a \emph{reference point} of translation between the two classes of coordinate systems, following the procedure outlined in \cite[Ch.~4]{coordinateconversion}.

\subsection{Channel Condition}
To model the channel condition for the \gls{ntn} propagation scenarios, we create the classes: 
\begin{itemize}
    \item \verb|ThreeGppNTNDenseUrbanChannelConditionMode|;
    \item \verb|ThreeGppNTNUrbanChannelConditionMode|;
    \item \verb|ThreeGppNTNSuburbanChannelConditionMode|; and
    \item \verb|ThreeGppNTNRuralChannelConditionMode|.
\end{itemize}
Each of these derives from the base class \texttt{Three\-Gpp\-Channel\-Condition\-Model}, which in turn implements the \texttt{Channel\-Condition\-Model} interface.

These channel condition classes interact with the remainder of the \texttt{spectrum} module as follows.
Whenever the \texttt{Get\-Channel\-Condition} method is called, the newly introduced NTN \texttt{Channel\-Condition\-Model} classes compute the channel state and cache it, along with its generation time. Then, the following calls to \texttt{Get\-Channel\-Condition} retrieve the previously stored value, if it has not expired. Otherwise, they compute a new \gls{los} condition. 

\subsection{Path Loss and Shadowing}
We implement the path loss and shadowing models of~\cite{38811} in four different classes, as depicted in Figure~\ref{fig:ntn-uml}. The latter extend the \texttt{Three\-Gpp\-Propagation\-Loss\-Model} class, which in turn implements the \texttt{Propagation\-Loss\-Model} interface.

The classes which implement this interface shall override the \texttt{Do\-Calc\-Rx\-Power}, returning the received power based on the positions of the communicating endpoints, and when considering frequency-flat phenomena only. 
In the case of the \gls{ntn} propagation scenarios of~\cite{38811}, these phenomena comprise the typical free space path loss, on top of tropospheric and ionospheric scintillation, shadow fading, clutter loss, and atmospheric absorption.

\subsection{Geocentric Mobility Models}
\label{sub:mobility}
Along with the geographic coordinate systems, we implement a new mobility model, i.e., \texttt{Geocentric\-Constant\-Position\-Mobility\-Model}, which allows ns-3 users to position nodes using real world coordinates.
Specifically, the latter class stores positions via the variable \verb|m_position|, which specifies their geographic coordinates.

When using these mobility models, the position of a node can be retrieved (set) using the methods \texttt{Get\-Geographic\-Postion} (\texttt{Set\-Geographic\-Postion}) and \texttt{Get\-Geocentric\-Position} (\texttt{Set\-Geocentric\-Position}). In turn, these methods rely on the functionality provided by the class \texttt{Geographic\-Position} for translating between different coordinate systems. 

Notably, the conversion from geocentric cartesian or geographic coordinates to the coordinate systems used by ns-3 \cite{ns3coordinates} uses by default the so-called reference point ``{Null Island}'' $(0,0,0)$. Nevertheless, ns-3 users are given the possibility of tuning this value by using the \texttt{Geocentric\-Constant\-Position\-Mobility\-Model} attribute \texttt{Set\-Coordinate\-Translation\-Reference\-Point}.

\subsection{Antenna Models}
The circular aperture reflector antenna model currently implemented in ns-3, i.e., \texttt{Parabolic\-Antenna\-Model}, is based on a parabolic approximation of the main lobe radiation pattern, as described in~\cite{parabolicantenna3gpp} and~\cite{parabolicantennamodel}. This simplification reduces the computational complexity of the field pattern calculation, by avoiding the Bessel functions evaluations that the circular aperture antenna would require, and using trigonometric approximations instead. 

As part of our contributions, we leverage the efficient implementation of the Bessel functions which has been introduced with C\texttt{++}17 to implement an exact circular aperture reflector antenna model. Specifically, we introduce this functionality extending the \texttt{Antenna\-Model} via the \texttt{Circular\-Aperture\-Antenna\-Model} class.
The latter allows ns-3 users to steer the pointing direction of the antenna via the \texttt{Set\-Orientation} and \texttt{Set\-Inclination} methods.
Similarly, the operating frequency and the aperture radius %
can be tuned by using the methods \texttt{Set\-Operating\-Frequency} and \texttt{Set\-Aperture\-Radius}.

\section{Examples and Comparisons}
\label{sec:results}
In this section we validate the accuracy of our ns-3 module for the NTN channel, and compare simulation results with the calibration reported in TR 38.821 \cite{38821}. Furthermore, we provide numerical results to measure link-level and end-to-end performance (including throughput and packet drop ratio). We focus on satellites, even though the model is valid for different NTN scenarios.

\subsection{Link-Level Results}
\label{sub:ll}
While ns-3 enables system-level simulations, an evaluation of the link-level performance is still useful to validate the technical accuracy of our module. Hence, in this section we run link-level simulations to compare the calibration results from the 3GPP~\cite{38821} with results from our module. 

The 3GPP identifies 30 calibration study cases~\cite[Table~ 6.1.1.1-9]{38821}, which include a combination of different satellite orbits, frequency bands, and antenna configurations for the ground terminal. Link-level calibration results are reported in~\cite[Table~6.1.1.2]{38821}, including results for the \gls{fspl}, atmospheric loss (AL) and scintillation loss (SL), and the \gls{cnr}. 
Specifically, the \gls{cnr} is calculated as described in \cite[Section~6.1.3.1]{38821}. 

\begin{table}
\caption{Link-level Comparison Between the 3GPP Calibration Results (``3GPP'') and those Obtained in Simulations (``Obtained''). Header Acronyms:  Free Space Path Loss (FSPL), Atmospheric Loss (AL), Scintillation Loss (SL), Carrier-to-Noise Ratio (CNR). All Values Are in dB}
\label{tab:ll-comp}
\vspace{-0.3cm}
\begin{tabular}{|c|l|l|l|l|l|l|}
\hline
\textbf{SC} & \textbf{Tx} & \textbf{Source} & \textbf{FSPL} & \textbf{AL} & \textbf{SL} & \textbf{CNR} \\ \hline
\multirow{2}{*}{1} & \multirow{2}{*}{DL} & 3GPP & 210.6 & 1.2 & 1.1 & 11.6 \\
 &  & Obtained & 210.6 & 1.4 & 1.1 & 11.3 \\ \hline
\multirow{2}{*}{1} & \multirow{2}{*}{UL} & 3GPP & 214.1 & 1.1 & 1.1 & 0.5 \\
 &  & Obtained & 214.2 & 1.4 & 1.1 & 0.1 \\ \hline
\multirow{2}{*}{6} & \multirow{2}{*}{DL} & 3GPP & 179.1 & 0.5 & 0.3 & 8.5 \\
 &  & Obtained & 179.9 & 0.5 & 0.3 & 8.6 \\ \hline
\multirow{2}{*}{6} & \multirow{2}{*}{UL} & 3GPP & 182.6 & 0.5 & 0.3 & 18.4 \\
 &  & Obtained & 182.6 & 0.5 & 0.3 & 18.4 \\ \hline
\multirow{2}{*}{9} & \multirow{2}{*}{DL} & 3GPP & 159.1 & 0.1 & 2.2 & 6.6 \\
 &  & Obtained & 159.1 & 0.0 & 2.2 & 6.7 \\ \hline
\multirow{2}{*}{9} & \multirow{2}{*}{UL} & 3GPP & 159.1 & 0.1 & 2.2 & 2.8 \\
 &  & Obtained & 159.1 & 0.0 & 2.2 & 2.4 \\ \hline
\multirow{2}{*}{14} & \multirow{2}{*}{DL} & 3GPP & 164.5 & 0.1 & 2.2 & 7.2 \\
 &  & Obtained & 164.5 & 0.0 & 2.2 & 7.3 \\ \hline
\multirow{2}{*}{14} & \multirow{2}{*}{UL} & 3GPP & 164.5 & 0.1 & 2.2 & -2.6 \\
 &  & Obtained & 164.5 & 0.0 & 2.2 & -3 \\ \hline
\end{tabular}
\end{table}

For this comparison we selected four study cases, considering both \gls{ul} and \gls{dl} transmissions, that illustrate four  representative NTN scenarios. Specifically: 
\begin{itemize}
    \item Study Case 1 (SC1): GEO satellite, 45 degrees of elevation, VSAT antenna for the ground terminal,	Ka-band.
    \item Study Case 6 (SC6): LEO satellite at 600 km, 90 degrees of elevation, VSAT antenna for the ground terminal,	Ka-band.
    \item Study Case 9 (SC9): LEO satellite at 600 km, 90 degrees of elevation, UPA antenna for the ground terminal,	S-band.
    \item Study Case 14 (SC14): LEO satellite at 1200 km, 90 degrees of elevation, UPA antenna for the ground terminal,	S-band.
\end{itemize}
The complete list of parameters used in the calibration can be found in \cite[Section 6.1]{38821}. 
In Table~\ref{tab:ll-comp} we report the calibration results (``3GPP'') and those from our simulations (``Obtained''). We can see that, despite some minor variations, numerical result are compatible under all metrics, thereby validating the accuracy of our module. 
As expected, the \gls{fspl} increases as the distance between the ground terminal and the satellite, as well as the carrier frequency, increase, due to the more severe effect of atmospheric losses. In particular, the impact of the carrier frequency is quite significant: for LEO satellites, for example, the \gls{fspl} grows from around 160 dB in the S-band (SC6) to around 180 dB in the Ka-band (SC9). In any case, we can see that,  even considering long-range GEO satellites in the Ka-band, the \gls{cnr} is large enough to support adequate levels of communication, especially in the downlink. 
We shed light on two main trends. First, uplink communication is generally worse than downlink, except for SC6. This is reasonable, and is due to the fact that ground terminals are more constrained in terms of power availability, capacity, and size (e.g., for antenna deployment).
Second, according to the 3GPP model, LEO satellites have more severe hardware constraints than GEO satellites (e.g., LEO's effective isotropic radiated power (EIRP) in the Ka-band is as low as 36 dBW, vs. 66 dBW for GEO): as a result, the CNR for SC6 is around 50\% lower than for SC1, which makes LEO communication more~difficult.

\subsection{Frequency Test}
\begin{figure}[t]
    \centering 
    \setlength\fwidth{0.85\columnwidth}
    \setlength\fheight{0.32\columnwidth}
    \input{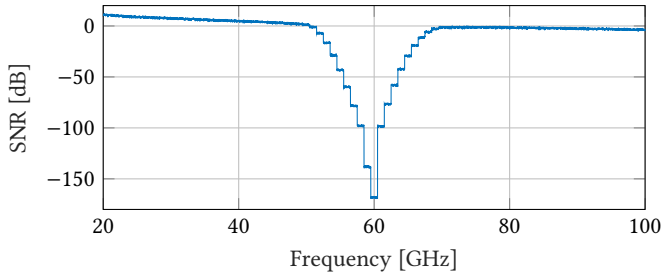}
    \caption{SNR vs Carrier Frequency. We Consider a GEO Satellite, with the Parameters in Table~\ref{tab:frequency-test}}
    \label{fig:frequency-test}
\end{figure}

While the 3GPP identifies the S-band (at 2 GHz) and the Ka-band (at 20 GHz for \gls{dl} and 30 GHz for \gls{ul}) as frequencies of interests, the NTN channel model is valid for a wide range of frequencies, from 0.5 GHz to 100 GHz. 
Therefore, in Figure~\ref{fig:frequency-test} we plot the \gls{snr}, which is an indication of the quality of the channel, as a function of the carrier frequency, with a resolution of 8 MHz. The other parameters are summarized in Table~\ref{tab:frequency-test}.

\begin{table}
\caption{Simulation Parameters}
\label{tab:frequency-test}
\begin{tabular}{|l|l|}
\hline
\textbf{Parameter} & \textbf{Value} \\ \hline
{Frequency} & {20 GHz $\div$ 100 GHz} \\\hline
{Satellite orbit} & {GEO} \\\hline
{Satellite altitude} & {35\,786 km} \\\hline
{Elevation angle} & {90 deg} \\\hline
{Transmission mode} & {Downlink} \\\hline
{Transmission power} & {37.5 dBm} \\\hline
{Satellite  antenna} & {Circular aperture (Gain: 58.5 dB)} \\\hline
{Terminal antenna} & {VSAT (Gain: 39.7 dB)} \\\hline
 {Scenario} & {Suburban} \\\hline
\end{tabular}
\end{table}

We observe that the \gls{snr} decreases linearly (in the log scale) as the frequency increases, with a severe degradation at 60 GHz. This is because of the impact of atmospheric absorption described in Section~\ref{sub:atm}, more specifically the additional signal attenuation experienced at 60 GHz due to oxygen, which can be as large as 15 dB/km.

\subsection{Mobility Test}
Our new mobility model for NTN (see Section~\ref{sub:mobility}) makes it possible to change the position of the nodes during the simulation, thus to evaluate the effect of different parameters such as the elevation angle, the antenna radiation pattern, and the altitude. 

\begin{figure}
    \centering 
    \setlength\fwidth{0.79\columnwidth}
    \setlength\fheight{0.32\columnwidth}
    \input{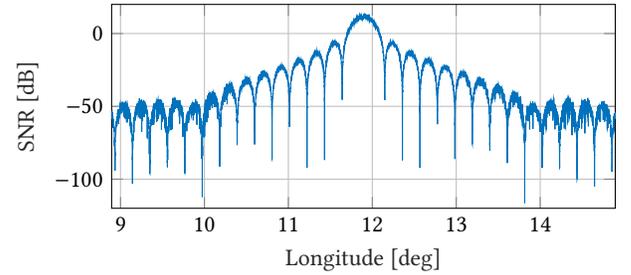}
    \caption{SNR vs Angular Positions of a GEO Satellite, with the Parameters in Table~\ref{tab:frequency-test}}
    \label{fig:movement-test}
\end{figure}

First, we run simulations where we iteratively change the coordinates of a satellite, which traces an arc of 6 degrees in the GEO orbit (from 8.8 to 14.8 degrees of longitude). %
The receiving node on the ground is deployed so that it is perpendicular to the satellite in the mid point of its trajectory. The rest of the parameters are set as in Table~\ref{tab:frequency-test}. Notably, the orientation of the antenna is not changed during the simulation, so that satellite and ground terminal are perfectly aligned only when the former is exactly perpendicular to the latter. 
In Figure~\ref{fig:movement-test} we plot the corresponding \gls{snr}, which resembles the circular antenna radiation pattern of the satellite as per Equation~\eqref{eq:eq4}, which therefore defines the power profile of the received signal.

\begin{figure}[t]
    \centering 
    \setlength\fwidth{0.85\columnwidth}
    \setlength\fheight{0.32\columnwidth}
    \input{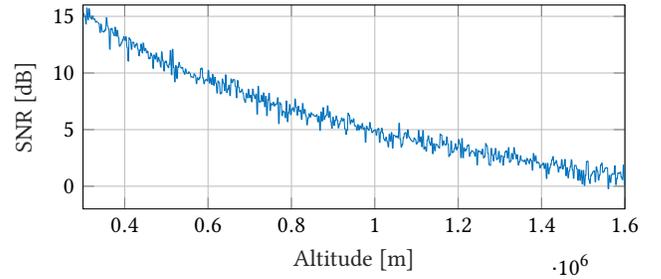}
    \caption{SNR vs Altitude of a LEO Satellite, with the Parameters in Table~\ref{tab:frequency-test}}
    \label{fig:altitude-test}
\end{figure}

Second, in Figure~\ref{fig:altitude-test} we plot the SNR for different  values of the altitude of the satellite, from 300 to 1\,600 km to consider different LEO satellite architectures. We see that the SNR decreases as the altitude of the satellite increases, even though it is consistently above 0 dB in all configurations.

\subsection{End-to-End Performance}
Unlike other simulators, ns-3 incorporates an accurate model of the whole ISO/OSI protocol stack, thus enabling scalable end-to-end simulations. Notably, end-to-end results can be collected to validate and measure the performance of communication networks, so ns-3 stands out as a valid tool to dimension NTN systems too.

\begin{table}
\caption{End-to-end Performance. Acronyms: Orbit Type (OT), Transmit Power (TxP), Throughput (TP), Drop Rate (DR), Frequency Band (FB), Terminal Antenna (TA)}
\label{tab:systemlev}
\begin{tabular}{|l|l|l|l|l|}
\hline
\textbf{SC} & \textbf{1} & \textbf{6} & \textbf{9} & \textbf{14} \\ \hline
\textbf{OT}       & GEO & \multicolumn{2}{c|}{LEO (600 km)} & LEO (1200 km) \\ \hline
\textbf{TxP}       & 37.52 dBm & 21.52 dBm & 48.77 dBm & 54.77 dBm \\ \hline
\textbf{FB} & Ka-band   & Ka-band   & S-band    & S-band    \\ \hline
\textbf{TA}    & VSAT      & VSAT      & UPA  & UPA  \\ \hline
\textbf{TP}     & 3.811 Mbit/s   & 3.286 Mbit/s      & 4.101 Mbit/s      & 5.161 Mbit/s        \\ \hline
\textbf{DR}      & 0.61      & 0.67      & 0.45      & 0.36      \\ \hline
\end{tabular}
\end{table}

To do so, we consider a downlink application transmitting data (in the form of UDP packets) at a constant rate of 10 Mbit/s. We test the four study cases presented in Section~\ref{sub:ll}, so as to consider both GEO and LEO satellites, and different altitudes, antenna configurations, and both communication in the Ka- and S-bands. 
Simulation results are reported in Table~\ref{tab:systemlev} in terms of end-to-end throughput (TP) and packet drop rate (DR), defined as one minus the ratio between the number of received packets and the total number of packets sent at the application layer.
We observe that both GEO-to-ground and LEO-to-ground communications are feasible, provided that the satellite operates with large-scale antennas offering fine-grained beams on the ground, and at high transmit power. 
Notice that the DR is quite significant, especially in the Ka-band, which requires the design of appropriate Automatic Repeat reQuest (ARQ) schemes to deal with retransmissions.

\balance

\section{Conclusions}
\label{sec:conclusions}
This paper presented an ns-3 implementation of the \gls{3gpp} channel model for NTN, developed following the specifications provided in \cite{38811}.
The code, which is publicly available at~\cite{ntngitlab}, integrates well with the rest of the ns-3 framework, and enables full-stack end-to-end simulations in different \gls{ntn} scenarios. After introducing the 3GPP NTN channel, we described in detail our ns-3 implementation, including simulation scenarios, new attenuation, fading, and mobility models for the space/air channel, and a new coordinate system specifically tailored to satellites. Finally, we run link-level and end-to-end simulations to validate the accuracy of our module, also against 3GPP calibration results reported in~\cite{38821}.

As part of our future work, we plan to further extend our NTN module to incorporate additional functionalities, for example a delay model able to accurately characterize the propagation time (thus to evaluate the performance of the higher layers, in particular the transport layer), the support for moving nodes in orbits, and the simulation of ground terminals in indoor scenarios.

\begin{acks}
This work was partially supported by the European Union under the Italian National Recovery and Resilience Plan (NRRP) of NextGenerationEU, partnership on “Telecommunications of the Future” (PE0000001 - program “RESTART”).
\end{acks}

\bibliographystyle{ACM-Reference-Format}
\bibliography{sample-base}

\end{document}